\documentclass[11pt,a4paper]{article}
\usepackage[english] {babel}
\usepackage[latin1]{inputenc}
\usepackage[T1]{fontenc}

\usepackage{amsmath, amsthm}
\usepackage{amsfonts,amssymb}
\usepackage{url}
\usepackage{graphicx}

\usepackage{latexsym}
\usepackage{bbm}
\usepackage{verbatim}

\newcommand{\ind}{{\mathbbm 1}}
\newcommand{\E}{{\mathbb E}}

\newcommand{\Var}{{\mathbb V}}

\newtheorem{Rem}{Remark}

\usepackage[margin=2.8cm]{geometry}
\begin{document}

\title{\bf Estimating with kernel smoothers the mean of functional data  in a finite population setting. A note on variance estimation in presence of partially observed trajectories}

\author{Herv\'e {\sc Cardot}$^{(a)}$, Anne  {\sc De Moliner}$^{(a,b)}$  and Camelia {\sc Goga}$^{(a)}$ \\
(a)  Institut de Math\'ematiques de Bourgogne, UMR 5584 CNRS, \\ Universit\'e de Bourgogne, France \\ %\\ 9 av. Alain Savary, 21078 DIJON, FRANCE \\
(b) EDF, R{\&}D, ICAME-SOAD, France% 1 av. du G\'en\'eral de Gaulle, \\ 92141 CLAMART, FRANCE
}

\maketitle

\begin{abstract}
This paper studies, in a survey sampling framework with unequal probability sampling designs, three nonparametric kernel estimators for the mean curve in  presence of discretized trajectories with  missing values.
Their pointwise  variances are approximated thanks to linearization techniques. 
\end{abstract}

\smallskip


%--------------------------------------------------------------------------

\section{Introduction}

In the next few years in France, tens of millions of smart meters will be deployed and will collect the individual load curves, i.e. electricity consumption time series, of residential and business customers at short time steps (probably half hours). This deployment will result in a huge increase in the amount of available data for energy suppliers such as EDF (Electricit\'e de France) and power grid managers. However, it may be complex and costly to stock and exploit such a large quantity of information, therefore it will be relevant to use sampling techniques to estimate load curves of specific customer groups (e.g. market segments, owners of a specific equipment or clients of an energy supplier). See Cardot and Josserand (2011) or Cardot \textit{et al.} (2013b) for preliminary studies.

\par

Unfortunately, data collection, like every mass process, may undergo technical problems at every point of the metering and collecting chain resulting in missing values. This problem is very similar to nonresponse in survey sampling: it deteriorates the accuracy of the estimators and may generate bias if the clients affected by missing values are different from the clients with complete curves. 

There is a large literature dealing with the inference in presence of missing values (see \textit{e.g.} S\"arndal \& Lundstr\"om (2005) and Haziza (2009) for reviews) but as far as we know the case in which collected data are curves has not been addressed yet.

\par
In this paper we will use functional data analysis methods, adapted to the sampling framework and to the presence of missing values, in order to take advantage of the specificities of our problematic that is to say the strong correlations between the consumptions at the various instants and  the smoothness of the curves. More precisely, we suggest to adapt, in an unequal probability sampling context, kernel estimation techniques that have initially been developed to deal with longitudinal data (Staniswalis and Lee, 1998) and sparse functional data (Hall \textit{et al.} 2006).
The asymptotic behavior of the Mean Square Error of the estimators (a very close estimator in fact) is given by Hart \& Wehrly (1986) and Hall \textit{et al.} (2006)  under the assumption that the number of measurements and the number of observations tends to infinity.
In a finite population setting with unequal probability sampling designs, the properties of local polynomial smoothers  with noisy measurements at at finite number of instants of time but without non response have been studied in Cardot \textit{et al.} (2013a).

The context we consider in this work is different and new. We suppose that some curves of the finite population under study are partially observed during periods that are random.   
The second section fixes notations and presents the three proposed kernel estimators of the mean load curve when pieces of trajectories are missing. In Section 3, the approximate variance of these estimators are derived. 
Note that our derivations are quite general and remain true for  linear smoothers (local polynomials, series expansion, smoothing splines).
The important particular case of stratified sampling is studied more precisely in Section 4 and we get that H\'ajek type estimators seem to be preferable, in this context, to weighted Horvitz-Thompson estimators. Finally some comments about estimation and choice of the tuning parameters are given in Section 5. Some technical details about kernels estimators are postponed in an Appendix.

%%%%%%%%%%%%%%%%%%%%%%%%%%%%%%%%%%%%%%%%%%%%%%%%
\section{The functional observations,  the non response mechanism  and the estimators}

We consider a population $U$,  with known size $N$, of (load) curves  defined over a time interval $[0,T]$:  for each unit $k$ in $U$, we have a function of time $Y_k(t), \ t \in [0,T]$, where  the continuous index  $t$ represents time.

The aim is to estimate the mean load curve $\mu$ (or the total trajectory = $N \mu$) over the population
\begin{align}
\mu(t) &= \frac{1}{N} \sum_{k \in U} Y_k(t), \quad  t \in [0,T],
\end{align}
when only a sample (drawn randomly from the population $U$) of the units is available and some parts of the sampled trajectories are missing. 
\subsection{Kernel smoothing of the mean trajectory}

With real data, the trajectories $Y_k$ are not observed at all instants $t \in [0,T]$ but at some discrete  time instants, $0\leq t_1< \ldots, t_j< \ldots< t_d \leq T$, which are supposed to be the same for all data and equispaced, so that $t_j = T(j-1)/(d-1)$. For example in Cardot and Josserand (2011),  the measurements are made every half an hour over a period of two weeks.

In this ideal discretized framework a simple continuous approximation to the  function $\mu$, denoted by  $\tilde{\mu}(t),$  can be obtained at all instants $t \in [0,T]$ by  applying a kernel smoother (see Staniswalis and Lee, 1998). For that, let us introduce a kernel $K(.)$, \textit{i.e} a continuous and positive function, symmetric around zero (see \textit{e.g.} Hart, 1997 for a more precise definition as well as examples). Classical kernels are the Gaussian kernel defined by $K(x) = \frac{1}{\sqrt{2 \pi}} \exp(-x^2/2)$ and  the Epanechnikov kernel which is defined by
$K(x) = \frac{3}{4} (1-x^2) \ind_{\{|x| \leq 1\}}$. For all instants $t \in [0,T]$, employing kernel smoothing leads to the following smooth approximation to $\mu(t)$,
\begin{align}
\tilde{\mu}(t) &=  \frac{ \displaystyle \sum_{k=1}^N \sum_{j=1}^{d} K\left(\frac{t-t_{j}}{h}\right) Y_k(t_{j})}{ \displaystyle \sum_{k=1}^N \sum_{j=1}^{d} K\left(\frac{t-t_{j}}{h}\right) }
\label{def:mutildeHa}
\end{align}
with a bandwidth $h$ whose aim is to control the smoothness of the approximation. Larger values of $h$ lead to smoother estimates, with a larger bias and a smaller variance whereas small values of $h$ lead to estimates that may have many oscillations, with a small bias but a larger variability.
When the design points $t_1, \ldots, t_d$ are the same for all the curves, expression~(\ref{def:mutildeHa}) can be simplified as follows,
\begin{align}
\tilde{\mu}(t) &= 
  \sum_{j=1}^{d} w(t,t_j,h) \mu(t_j),
  \label{def:mutildeHT}
\end{align}
with smoothing weights
\[
w(t,t_j,h) =  \frac{K\left(\frac{t-t_{j}}{h}\right)}{\sum_{j=1}^d K\left(\frac{t-t_{j}}{h}\right)},
\]
so that $\tilde{\mu}(t)$ is simply obtained by smoothing  the discretized population mean trajectory, $(\mu(t_1), \ldots, \mu(t_d))$.

\subsection{Sampling designs and kernel estimators with full response}

It is assumed  now that only a part the population $U$ is observed and we  denote by $s \subset U$ a sample drawn randomly from $U$, with fixed size $n$.  
For $k$ and $\ell  \in \{1, \ldots, N\}$, we denote by $\pi_k$ and $\pi_{k\ell}$ the first and second order inclusion probabilities: $\pi_k = \mathbb{P}\left[k \in s \right]$ and $\pi_{k\ell} = \mathbb{P}\left[ k \in s \ \& \ \ell \in s \right]$. These inclusion probabilities are supposed to be strictly positive.

With full response, the Horvitz-Thompson (Horvitz-Thompson, 1952) estimator of the smooth approximation $\tilde{\mu}(t)$ is simply obtained by replacing the $\mu(t_j)$'s in (\ref{def:mutildeHT}) by their Horvitz-Thompson estimations $\widehat{\mu}(t_j)$:
\begin{align}
\widehat{\mu}_{HT}(t_j) & = \frac{1}{N} \sum_{k \in s} \frac{Y_k(t_j)}{\pi_k},
\end{align}
so that we get, for $t  \in [0,T]$, 
\begin{align}
\widehat{\mu}_{HT}(t) & = 
  \sum_{j=1}^{d} w(t,t_j,h) \ \widehat{\mu}(t_j).
\label{def:HTwnor}  
\end{align}
Another estimator of $\tilde{\mu}(t)$ can be defined by considering a ratio or H\'ajek (H\'ajek, 1971) point of view. The estimator is obtained by replacing the $\mu(t_j)$'s in (\ref{def:mutildeHT}) by their H\'ajek estimator:
\begin{align}
\widehat{\mu}_{Ha}(t_j)&=\frac{ \sum_{k \in s} \frac{Y_k(t_{j})}{\pi_k}}{\sum_{k \in s} \frac{1}{\pi_k}}.
\label{def:Hajektj}
\end{align}
We get
\begin{align}
\widehat{\mu}_{Ha}(t) & =\sum_{j = 1}^d w(t,t_j,h) \widehat{\mu}_{Ha}(t_j)
\label{def:Ha2wnor}\\
&= \frac{ \displaystyle \sum_{k \in s} \sum_{j=1}^{d} K\left(\frac{t-t_{j}}{h}\right) \frac{Y_k(t_{j})}{\pi_k}}{ \displaystyle \sum_{k \in s} \sum_{j=1}^{d} K\left(\frac{t-t_{j}}{h}\right) \frac{1}{\pi_k}}.
\label{def:Ha1wnor}
\end{align}
The H\'ajek-type estimator  is in fact a regression estimator which takes into account auxiliary information (S\"arndal \textit{et al.}, 1992). The estimator defined in (\ref{def:Hajektj}) can be obtained by considering a one-way ANOVA model $\xi$ for $Y_k$ at each instant $t_j:$  $E_{\xi}(Y_k(t_j))=\beta_j(t_j)$ and $Var_{\xi}(Y_k(t_j))=\sigma_j^2(t_j)$ for all $k\in U,$ while (\ref{def:Ha1wnor}) can also be obtained by considering a one-way ANOVA model $\xi$  but for the smoothed trajectory $\tilde Y_k(t)=\sum_{j=1}^dw(t,t_j,h)Y_k(t_j)$. These models can be seen as particular cases of the functional linear models considered in  Cardot \textit{et al. }~(2013c).

\subsection{Nonparametric estimators with non response}

The individual trajectories of the sample $s$ are not always observed at all the discretization points and some parts may be missing. 
To take account of the non response mechanism we introduce a  response random variable defined as follows. We define  the continuous time process  $r_{k}(t)$ which takes value 1 if $Y_k$ can be observed at instant $t$ and 0 else. This binary continuous time stochastic process is supposed to be independent of the values of the trajectories as well as of the sampling design. Nonetheless, the response probability  is allowed to  depend on $k$ and on time. 
For each unit $k$ in the population, we denote the probability of response at instant $t_j$  by 
\begin{align*}
\theta_k(t_j) &= \mathbb{P}\left[ r_k(t_j)=1\right],
\end{align*}
 and the probability of response at both instants $t_j$ and $t_{j'}$ by 
\begin{align*}
\theta_k(t_j,t_{j'}) = \mathbb{P}\left[ r_k(t_j)=1 \ \& \ r_k(t_{j'})=1\right].
\end{align*}
Note that, for simpler calculus, we could also assume that there is a small number of response homogeneity groups (see  for example S\"arndal \& Lundstr\"om, 2005), that is to say that all elements within the same group respond with the same probability, and in an independent manner. 

Taking now the non response mechanism into account, we can consider  three different estimators of  $\tilde{\mu}(t)$ based on reweighting and smoothing. A first one, derived from (\ref{def:HTwnor}), is a smoothed Horvitz-Thompson estimator that takes account of non response, 
\begin{align}
\widehat{\mu}_{r,HT}(t) &=    \frac{1}{N} \sum_{j=1}^{d} w(t,t_j,h) \left( \sum_{k \in s} \frac{r_k(t_j)Y_k(t_j)}{\theta_k(t_j) \pi_k} \right).
\label{def:muhatrHT}
\end{align}
The second one derived from (\ref{def:Ha1wnor}), can be seen as a H\'ajek type estimator of two smoothed estimators,
\begin{align}
\widehat{\mu}^{(1)}_{r,Ha}(t) &= \frac{\displaystyle  \sum_{j=1}^{d} K(h^{-1}(t-t_{j})) \left( \sum_{k \in s}    \frac{r_k(t_j)}{\theta_k(t_j)} \frac{Y_k(t_{j})}{\pi_k} \right)}{\displaystyle \sum_{j=1}^{d} K(h^{-1}(t-t_{j})) \left( \sum_{k \in s}   \frac{r_k(t_j)}{\theta_k(t_j)} \frac{1}{\pi_k}\right)}.
\label{def:muhatr}
\end{align}
The last estimator  can also be seen as a smoothed H\'ajek estimator and is derived from (\ref{def:Ha2wnor}). It is defined, for $t \in [0,T],$ by
\begin{align}
 \widehat{\mu}^{(2)}_{r,Ha}(t) 
 %&=  \sum_{j=1}^dw(t,t_j,h) \ \frac{ \displaystyle  \sum_{k \in s}\frac{r_k(t_j)}{\theta_k(t_j)} \frac{Y_k(t_{j})}{\pi_k} }{ \displaystyle  \sum_{k \in s}   \frac{r_k(t_j)}{\theta_k(t_j)} \frac{1}{\pi_k}} \nonumber \\
&=  \sum_{j=1}^dw(t,t_j,h) \ \frac{ \widehat{Y}(t_j)}{\widehat{N}(t_j)}, 
 \label{def:muhatrv2} 
\end{align} 
where $\widehat{Y}(t_j) =\sum_{k \in s}\frac{r_k(t_j)}{\theta_k(t_j)} \frac{Y_k(t_{j})}{\pi_k}$ and $\widehat{N}(t_j)=\sum_{k \in s}   \frac{r_k(t_j)}{\theta_k(t_j)} \frac{1}{\pi_k}$.
Note that, since the response probability is different from one instant to another, we have now two different H\'ajek-type estimators  $\widehat{\mu}^{(1)}_{r,Ha}(t)$ and $\widehat{\mu}^{(2)}_{r,Ha}(t),$ while with full response they are the same.

%%%%%%%%%%%%%%%%%%%%%%%%%
\section{Variance of the estimators}

We first show, under general conditions,  that the approximation error and the bias are negligible compared to the variance. This explains why we focus on variance estimation of the three proposed estimators in the presence of non response.

We denote in the following  by $\mathbb{E}_p$ the expectation with respect to the sampling design and by $\mathbb{E}_R$ the expectation with respect to the non response mechanism. When there is no subscript, the expectation $\mathbb{E}$ is considered both with respect to the sampling design and the non response random mechanism.

\subsection{Approximation error and bias}

We first consider the Horvitz-Thompson estimator $\widehat{\mu}_{r,HT}(t)$ defined in (\ref{def:muhatrHT}) and we clearly have that 
\begin{align}
\mathbb{E}(\widehat{\mu}_{r,HT}(t)) &= \sum_{j=1}^d w(t,t_j,h) \mathbb{E} \left[ \sum_{k \in s} \frac{r_k(t_j)Y_k(t_j)}{\theta_k(t_j) \pi_k} \right] \nonumber \\
 & = \sum_{j=1}^d w(t,t_j,h) \mu(t_j) \nonumber \\
 & = \tilde{\mu}(t)
\end{align}
so that it is unbiased for $\tilde{\mu}(t)$. Thus, the mean square error satisfies
\begin{align}
\E \left[ \widehat{\mu}_{r,HT}(t) - \mu(t) \right]^2 & =  \left|  \tilde{\mu}(t) - \mu(t) \right| ^2 +  \Var \left( \widehat{\mu}_{r,HT}(t) \right).
\label{def:eqmHTr}
\end{align}

Furthermore, we can show under general regularity conditions  on the mean trajectory given in the Appendix  and if, as the population size $N$ tends to infinity, the bandwidth $h$ tends to zero and the number of design points tends to infinity, satisfying  $2h >  (d-1)^{-1}$, that  the approximation error  is bounded, for some constant $C_t$, as follows
\begin{align}
 \left|  \tilde{\mu}(t) - \mu(t) \right| &\leq C_t h^\beta.
\label{approx:errormutilde} 
\end{align}

Combining (\ref{def:eqmHTr}) and (\ref{approx:errormutilde}), this means that, provided that $\sqrt{n} h^{\beta} \to 0$ as the sample size $n$ tends to infinity,   the approximation error, $\widetilde{\mu}(t) -\mu(t)$, is negligible  compared to the sampling error, which tends to zero in probability at most at rate $1/\sqrt{n}$. 
%and $\E \left[ \widehat{\mu}_{r,HT}(t) - \mu(t) \right]^2 \approx \Var \left( \widehat{\mu}_{r,HT}(t) \right)$. 
Note that this condition on the bandwidth $h$, which must be small, and the sample size also implies that the grid of discretization points must be dense enough so that  $\sqrt{n} \max_{j}|t_{j+1}-t_j|^\beta \to 0$.  In that case, the mean square error of the Horvitz-Thompson estimator can be approximated by its variance,  
\[
\E \left[\widehat{\mu}_{r,HT}(t) - \mu(t) \right]^2 \approx \Var \left( \widehat{\mu}_{r,HT}(t) \right).
\]

\begin{Rem}
The condition $\sqrt{n} \max_{j}|t_{j+1}-t_j|^\beta \to 0$, which can be written $n d^{-2\beta}=o(1)$ is important to decide if the approximation error can be neglected or not. If for example the mean trajectory $\mu$ is Lipschitz ({\it i.e.} $\beta=1$) then the number $d$ of discretization points is not required to be very large compared to the sample size $n$ since the condition can be written $n^2/d =o(1)$. On the other hand, if the mean trajectory is rough ($\beta$ small) then the bias is negligible only when the number $d$ is large (and potentially very large if $\beta$ is close to zero). In this case, smoothing over time with kernels is certainly not be the best strategy to estimate the mean at each point $t_j$ and a  parametric point of view, instant by instant, may give better results. 
\end{Rem}

The estimators $\widehat{\mu}^{(1)}_{r,Ha}(t)$ and $\widehat{\mu}^{(2)}_{r,Ha}(t)$ are not unbiased estimators of $\tilde{\mu}(t).$ They are ratio estimators with unbiased estimators of each component (numerator and denominator) of the ratio. Nevertheless, they are asymptotically unbiased for $\tilde{\mu}(t)$ and under previous conditions, their squared biais plus their squared approximation error are negligible compared to their variance. This means that their mean square errors can also be approximated by their variances.

\subsection{Variance of the Horvitz-Thompson estimator}

The derivations made below are greatly inspired by the Chapter 15 of S\"arndal \textit{et al.} (1992).
Let us first decompose the variance of $\widehat{\mu}_{r,HT}(t)$  by using the classical formula:
\begin{align}
\Var( \widehat{\mu}_{r,HT}(t)) &= \Var_R\E_p \left( \widehat{\mu}_{r,HT}(t) - \tilde{\mu}(t) |s_r \right)+\E_R\Var_p \left( \widehat{\mu}_{r,HT}(t) - \tilde{\mu}(t) |s_r \right),
\label{var:HTr}
\end{align}
where $s_r$ is the set of samples of respondents at each instant  $\{t_1, \ldots, t_p\}$. For simpler notations and shorter equations, we denote $w(t,t_j,h)$ by $w_j(t)$.
We have
\begin{align*}
\E_p( \widehat{\mu}_{r,HT}(t) - \tilde{\mu}(t) |s_r) & = \frac{1}{N} \sum_{j=1}^{d} w_j(t) \left[ \sum_{k \in U} Y_k(t_j)\left( \frac{r_k(t_j)}{\theta_k(t_j)} -1 \right)\right]
\end{align*}
and, by independence between units in the non response mechanism,
\begin{align}
\Var_R\E_p( \widehat{\mu}_{r,HT}(t) - \tilde{\mu}(t) |s_r) &= \frac{1}{N^2}  \sum_{k \in U} \sum_{j=1}^{d} w_j^2(t) Y_k^2(t_j) \frac{1- \theta_k(t_j)}{\theta_k(t_j)}  \nonumber \\ 
&  +\frac{1}{N^2}  \sum_{k \in U} \sum_{j=1}^{d}\sum_{j' \neq j} w_j(t) w_{j'}(t) Y_k(t_j) Y_k(t_{j'}) \frac{ \theta_k(t_j,t_{j'}) - \theta_k({t_j})\theta_k(t_{j'})}{\theta_k({t_j})\theta_k(t_{j'})}
\label{var:HTr1}
\end{align}
Define $\Delta_{kl}=\pi_{kl}-\pi_{k}\pi_{l}$ for $k \neq l$ and $\Delta_{kk} = \pi_k(1-\pi_k)$.
We have
\begin{align*}
\Var_p( \widehat{\mu}_{r,HT}(t) - \tilde{\mu}(t) |s_r) &= \frac{1}{N^2}  \sum_{k,l \in U} \frac{\Delta_{kl}}{\pi_k \pi_l} \left( \sum_{j=1}^d w_j(t) \frac{r_l(t_j)}{\theta_l(t_j)}Y_l(t_j) \right)\left( \sum_{j'=1}^d w_{j'}(t) \frac{r_k(t_{j'})}{\theta_k(t_{j'})}Y_k(t_{j'}) \right) 
\end{align*}
and taking the expectation with respect to the non response mechanism, we get
\begin{align}
\E_R\Var_p( \widehat{\mu}_{r,HT}(t) - \tilde{\mu}(t) |s_r) &=  \frac{1}{N^2}  \sum_{k\in U} \frac{1-\pi_k}{\pi_k}  \sum_{j=1}^d\sum_{j',j'\neq j=1}^d w_j(t)  w_{j'}(t) Y_k(t_j)Y_k(t_j')\frac{\theta_k(t_j,t_{j'})}{\theta_k(t_j) \theta_k(t_{j'})} \nonumber \\
&  + \frac{1}{N^2} \sum_{k\in U} \frac{1-\pi_k}{\pi_k} \sum_{j=1}^d w_j^2(t)  Y_k^2(t_j) \frac{1}{\theta_k(t_j)} \nonumber \\
 &+ \frac{1}{N^2}  \sum_{k\in U}\sum_{l \neq k\in U}  \frac{\Delta_{kl}}{\pi_k\pi_l}  \left(\sum_{j=1}^d w_j(t)  Y_k(t_j)\right)\left(\sum_{j'=1}^d w_{j'}(t)  Y_l(t_{j'})\right) 
\label{var:HTr2}
\end{align}
Combining (\ref{var:HTr1}) and (\ref{var:HTr2}) in (\ref{var:HTr}), we get, after some algebra, the following expression for the variance of $\widehat{\mu}_{r,HT}(t)$, at each instant $t$ in $[0,T]$,
\begin{align}
\Var( \widehat{\mu}_{r,HT}(t)) &= \frac{1}{N^2}  \sum_{k,l\in U}  \frac{\Delta_{kl}}{\pi_k\pi_l}  \left(\sum_{j=1}^d w_j(t)  Y_k(t_j)\right)\left(\sum_{j'=1}^d w_{j'}(t)  Y_l(t_{j'})\right) \label{var:HTrv1}\\
&+ \frac{1}{N^2}  \sum_{k\in U} \frac{1}{\pi_k}  \sum_{j,j'=1}^d w_j(t)  w_{j'}(t) Y_k(t_j)Y_k(t_{j'})\frac{\theta_k(t_j,t_{j'}) - \theta_k(t_j) \theta_k(t_{j'})}{\theta_k(t_j) \theta_k(t_{j'})} \label{var:HTrv2} 
\end{align}
with the convention  that $\theta_k(t_j,t_{j}) = \theta_k(t_j)$. The part of the variance given in (\ref{var:HTrv1}) corresponds to the sampling variance whereas the additional variance term in (\ref{var:HTrv2}) is due to the non response. \medskip

\subsection{Variance approximation for the H\'ajek estimators}

The variance of the estimator $\widehat{\mu}^{(1)}_{r,Ha}(t)$ defined in (\ref{def:muhatr}) can be approximated thanks to a linearization technique  (see Deville, 1999) with respect to the sampling distribution and the non response mechanism. 
Indeed, it is a ratio of two linear estimators whose expressions are similar to the expression of $\widehat{\mu}_{r,HT}(t)$. We have
\begin{align}
\Var \left( \widehat{\mu}^{(1)}_{r,Ha}(t) \right)& \approx \Var \left( \sum_{k \in s} \sum_{j=1}^d \frac{r_k(t_j)}{\theta_k(t_j)} \frac{u_{kj}^{(1)}(t)}{\pi_k} \right) 
\label{var:apprxha1}
\end{align}
where the ``linearized'' variable $u^{(1)}_{kj}(t)$ of $\widehat{\mu}^{(1)}_{r,Ha}(t)$ is defined as follows
%\begin{align}
%u_{kj}(t) &=\frac{K\left(h^{-1}(t-t_{j})\right)}{N \sum_{j=1}^{d} K\left(h^{-1}(t-t_{j})\right)}\left(Y_{k}(t_j)-\tilde \mu(t) \right).
%\end{align}
\begin{align}
u_{kj}^{(1)}(t) &=\frac{1}{N}  w_j(t) \left(Y_{k}(t_j)-\tilde \mu(t) \right).
\end{align}

%with the notations $\check{Y}_{kj}(t) = K(h^{-1}(t-t_j))Y_k(t_j)$ et $\check{X}_{j} (t)= K(h^{-1}(t-t_j))$.  
%$\displaystyle u_{kj}(t) =\frac{1}{\sum_{k \in U} \sum_{j=1}^p  \check{X}_{j}(t)}\left(\check{Y}_{kj}(t)-\tilde \mu(t)\check{X}_{j}(t)\right)$, $\widetilde u_k(t)=\sum_{j=1}^pu_{kj}(t)$.  

\noindent After some algebra, we get the following expression for the variance:
\begin{align}
\Var \left( \widehat{\mu}^{(1)}_{r,Ha}(t) \right)& \approx \sum_{k\in U}\sum_{l\in U} \frac{\Delta_{kl}}{\pi_k\pi_l}\widetilde{u}_k^{(1)}(t)\widetilde{u}_l^{(1)}(t)+\sum_{k\in U}\frac{1}{\pi_k}\sum_{j,j'=1}^d\frac{u_{kj}^{(1)}(t)u_{kj'}^{(1)}(t)}{\theta_k(t_j)\theta_k(t_{j'})}(\theta_k(t_{j},t_{j'})-\theta_k(t_j)\theta_k(t_{j'}))\label{var:apprxha1.2}
\end{align}
where $\widetilde u_k^{(1)}(t)$ is the smoothed linearized variable trajectory, 
\[
\widetilde u_k^{(1)}(t) \ = \ \sum_{j=1}^du_{kj}^{(1)}(t)=\frac{1}{N}\sum_{j=1}^d  w_j(t) \left(Y_{k}(t_j)-\tilde \mu(t) \right).  
\]

\noindent For the third estimator, $\widehat{\mu}^{(2)}_{r,Ha}(t)$ defined in (\ref{def:muhatrv2}), we have
\begin{align}
\Var \left( \widehat{\mu}^{(2)}_{r,Ha}(t) \right)&= \sum_{j,j'=1}^d w_j(t) w_{j'}(t) \ \mathbb{C}ov \left( \frac{ \widehat{Y}(t_j)}{\widehat{N}(t_j)},\frac{ \widehat{Y}(t_{j'})}{\widehat{N}(t_{j'})} \right). 
\end{align}
Employing again a linearization technique, we have 
\begin{align*}
\mathbb{C}ov \left( \frac{ \widehat{Y}(t_j)}{\widehat{N}(t_j)},\frac{ \widehat{Y}(t_{j'})}{\widehat{N}(t_{j'})} \right)  & \approx \mathbb{C}ov \left( \frac{1}{N} \sum_{k \in s} \frac{r_k(t_j)}{\theta_k(t_j)} \frac{( Y_k(t_j) - \mu(t_j))}{\pi_k}, \frac{1}{N} \sum_{k \in s} \frac{r_k(t_{j'})}{\theta_k(t_{j'})} \frac{( Y_k(t_{j'}) - \mu(t_{j'}))}{\pi_k} \right),
\end{align*}
so that
\begin{align}
\Var \left( \widehat{\mu}^{(2)}_{r,Ha}(t) \right)& \approx \Var \left( \sum_{j=1}^d w_j(t) \frac{1}{N} \sum_{k \in s} \frac{r_k(t_j)}{\theta_k(t_j)} \frac{( Y_k(t_j) - \mu(t_j))}{\pi_k} \right) \nonumber \\
&=\Var \left( \sum_{k \in s} \sum_{j=1}^d  \frac{r_k(t_j)}{\theta_k(t_j)} \frac{u_{kj}^{(2)}}{\pi_k} \right),
 \label{var:apprxha2}
\end{align}
where 
$$
u_{kj}^{(2)}(t)=\frac{1}{N}w_j(t)  ( Y_k(t_j) - \mu(t_j))
$$
is the linearized variable of $\widehat{\mu}^{(2)}_{r,Ha}(t). $ 
A direct comparison of (\ref{var:apprxha2}) with (\ref{var:apprxha1}) gives us that the approximated  variance  of $\widehat{\mu}^{(2)}_{r,Ha}(t),$ which is based on linearization, is very similar to the approximated variance of $\widehat{\mu}^{(1)}_{r,Ha}(t)$. We have
% and the only difference is that $\tilde{\mu}(t_j)$ is replaced by $\mu(t_j)$:
\begin{align}
\Var \left( \widehat{\mu}^{(2)}_{r,Ha}(t) \right)& \approx \sum_{k\in U}\sum_{l\in U} \frac{\Delta_{kl}}{\pi_k\pi_l}\widetilde u_k^{(2)}(t)\widetilde u_l^{(2)}(t)+\sum_{k\in U}\frac{1}{\pi_k}\sum_{j,j'=1}^d\frac{u_{kj}^{(2)}(t)u_{kj'}^{(2)}(t)}{\theta_k(t_j)\theta_k(t_{j'})}(\theta_k(t_{j},t_{j'})-\theta_k(t_j)\theta_k(t_{j'}))\label{var:apprxha2.2}
\end{align}
%We can also write 
%\begin{eqnarray*}
%\widehat{\mu}^{(2)}_{r,Ha}(t)-\tilde{\mu}(t)=\sum_{j=1}^d w_j(t)\left(\frac{\hat Y(t_j)}{\hat N(t_j)}-\mu(t_j)\right)\simeq \sum_{j=1}^d w_j(t)\left(\sum_{k\in s}\frac{r_k(t_j)u_{kj}^{(2)}}{\theta_k(t_j)\pi_k}-\sum_{k\in U}u_{kj}^{(2)}\right),
%\end{eqnarray*}
where  $\widetilde u_k^{(2)}(t)$ is again a kind of smoothed linearized variable, this time of $u_{kj}^{(2)}(t):$
$$
\widetilde u_k^{(2)}(t) \ = \ \sum_{j=1}^du_{kj}^{(2)}(t)=\frac{1}{N}\sum_{j=1}^d  w_j(t) \left(Y_{k}(t_j)-\mu(t_j) \right).
$$
Using the fact that  $\sum_{j=1}^d w_j(t)=1,$ we obtain   
$$
\widetilde u_k^{(2)}(t) \ = \ \frac{1}{N}\sum_{j=1}^d  w_j(t) \left(Y_{k}(t_j)-\widetilde{\mu}(t) \right) \ = \ \widetilde u_k^{(1)}(t).
$$
This means that the first terms from the variances given in (\ref{var:apprxha1.2}) and (\ref{var:apprxha2.2}) are equal, confirming the fact that the estimators $\widehat{\mu}^{(1)}_{r,Ha}(t)$ and $\widehat{\mu}^{(2)}_{r,Ha}(t)$ are the same with full response.

\section{The particular case of stratified sampling with homogeneous response groups}

We consider now the  important  case of stratified sampling. 
The population $U$ is divided into $\Lambda$ strata, $U_\lambda, \lambda=1, \ldots, \Lambda$, with size $N_\lambda,$ so that 
 $U = \bigcup_{\lambda=1}^\Lambda U_\lambda$,  $U_\lambda \bigcap U_\ell = \emptyset$ if  $\lambda \neq \ell$ and $N= \sum_{\lambda=1}^\Lambda N_\lambda$. The  $\Lambda$ strata are built thanks to auxiliary information that is relevant to model the shape of the individual trajectories. \\
 The mean trajectory can be written
\begin{align}
\mu & =  \sum_{\lambda=1}^\Lambda \frac{N_\lambda}{N} \mu_\lambda 
\end{align}
where, for each $\lambda \in \{1, \ldots, \Lambda\}$, $\mu_\lambda$ is the mean trajectory in subpopulation $U_\lambda$,
\begin{align}
\mu_\lambda &= \frac{1}{N_\lambda} \sum_{k \in U_\lambda} Y_k. 
\end{align}

Different kernel smoothers may be considered in each stratum and the overall kernel approximation to $\mu$ is obtained by linear combination
\begin{align}
\widetilde{\mu} & =  \sum_{\lambda=1}^\Lambda \frac{N_\lambda}{N} \widetilde{\mu}_\lambda 
\end{align}
where, for each $\lambda$ and each $t \in [0,T]$, 
\begin{align}
\widetilde{\mu}_\lambda(t)  &=\sum_{j=1}^{d} w(t,t_j,h) \mu_\lambda(t_j),
\end{align}
where the smoothing weights $w(t,t_j,h)$ are defined as in (\ref{def:mutildeHT}) and are allowed to be different from one stratum to another.

\subsection{The estimators of the mean trajectory}
In each strata $\lambda$, a sample $s_\lambda$ of size $n_\lambda$ is drawn with simple random sampling without replacement. We suppose that the homogeneous response groups coincide with  the strata and  the non response mechanism is homogeneous within each stratum, so that the response probabilities are the same for all units within the stratum : if $k \in U_\lambda$, then $\mathbb{P}(r_k(t_j) = 1) = \theta_\lambda(t_j)$.
Using the fact that $\pi_k=n_{\lambda}/N_{\lambda}$ for all $k\in U_{\lambda},$ the Horvitz-Thompson estimator given in (\ref{def:muhatrHT}) becomes:
\begin{align*}
\widehat{\mu}_{HT}(t) &= \sum_{\lambda=1}^\Lambda \frac{N_\lambda}{N} \  \widehat{\mu}_{\lambda,HT}(t),
\end{align*}
where, for each $\lambda$,  
\begin{align}
\widehat{\mu}_{\lambda,HT}(t)  &=\sum_{j=1}^{d} w(t,t_j,h) \left( \frac{1}{n_\lambda} \sum_{k \in s_\lambda} Y_k(t_j)\frac{r_k(t_j)}{\theta_\lambda(t_j)}\right), \quad t \in [0,T]
\label{def:stratrHT}
\end{align}
is the Horvitz-Thompson estimator, based on the sample $s_{\lambda},$  of $\widetilde{\mu}_\lambda(t),$ the smoothed mean trajectory in subpopulation $U_\lambda.$ Moreover, $\frac{1}{n_\lambda} \sum_{k \in s_\lambda} Y_k(t_j)\frac{r_k(t_j)}{\theta_\lambda(t_j)}$ is the Horvitz-Thompson estimator of $\mu_\lambda.$

\noindent Now, the stratified sample $s=\cup_{\lambda=1}^{\Lambda} s_{\lambda}$ is poststratified (S\'arndal \textit{et al.}, 1992) and the poststrata coincide with the strata $U_\lambda. $ Then, the H\'ajek estimators $\widehat{\mu}^{(1)}_{\lambda,Ha}(t)$ and $\widehat{\mu}^{(2)}_{\lambda,Ha}(t)$ can also be written as 
\begin{align*}
\widehat{\mu}_{Ha}(t) &= \sum_{\lambda=1}^\Lambda \frac{N_\lambda}{N} \  \widehat{\mu}_{\lambda,Ha}(t),
\end{align*}
where, for each $\lambda$ and each $t \in [0,T],$ $\widehat{\mu}_{\lambda,Ha}(t)$ is either the estimator:
 \begin{align}
\widehat{\mu}^{(1)}_{\lambda,Ha}(t)  &=\frac{ \displaystyle \sum_{j=1}^{d} K(h^{-1}(t-t_{j}))  \left( \frac{1}{n_\lambda} \sum_{k \in s_\lambda} Y_k(t_j)\frac{r_k(t_j)}{\theta_\lambda(t_j)}\right)}{ \displaystyle \sum_{j=1}^{d} K(h^{-1}(t-t_{j}))  \left( \frac{1}{n_\lambda} \sum_{k \in s_\lambda} \frac{r_k(t_j)}{\theta_\lambda(t_j)}\right)},
\label{def:stratrHa}
\end{align}
or
\begin{align*}
\widehat{\mu}^{(2)}_{\lambda,Ha}(t)  &=\sum_{j=1}^{d} w(t,t_j,h) \left( \frac{ \sum_{k \in s_\lambda} Y_k(t_j)\frac{r_k(t_j)}{\theta_\lambda(t_j)}}{ \sum_{k \in s_\lambda} \frac{r_k(t_j)}{\theta_\lambda(t_j)}} \right)=\sum_{j=1}^{d} w(t,t_j,h) \left( \frac{ \sum_{k \in s_\lambda} Y_k(t_j)r_k(t_j)}{ \sum_{k \in s_\lambda} r_k(t_j)} \right). 
%\label{def:stratrHa2}
\end{align*}
We can remark that the response probabilities, $\theta_\lambda(t_j), j=1, \ldots, d$ have disappeared in the expression of $\widehat{\mu}^{(2)}_{\lambda,Ha}(t)$. As it is mentioned by S\"arndal \textit{et al.} (1992, Chapter 15),  this corresponds to ``doing nothing about the nonresponse, in the sense that the response model implies no difference in the weighting of respondent values $Y_k$''. 

\subsection{Variance formula for stratified sampling}

Note that by independence of the samples $s_1, \ldots, s_\Lambda$, we have that
\begin{align*}
\Var(\widehat{\mu}(t)) &= \frac{1}{N^2} \sum_{\lambda=1}^\Lambda  N_\lambda^2 \  \Var(\widehat{\mu}_{\lambda}(t)). 
\end{align*}
When considering the estimator $\widehat{\mu}_{\lambda,HT}(t)$ for $\widetilde{\mu}$, we get 
\begin{align*}
\Var(\widehat{\mu}_{\lambda,HT}(t)) &= \left( 1 - \frac{n_\lambda}{N_\lambda} \right) \frac{1}{n_\lambda} \frac{1}{N_\lambda - 1} \sum_{k \in U_\lambda} \left(  \widetilde{Y}_k(t) - \widetilde{\mu}_\lambda(t) \right)^2 \nonumber \\
&+ \frac{N_\lambda}{n_\lambda}\sum_{k\in U_\lambda}
\sum_{j,j'}^d w_j(t)  w_{j'}(t) Y_k(t_j)Y_k(t_{j'})\frac{\theta_\lambda(t_j,t_{j'}) - \theta_\lambda(t_j) \theta_\lambda(t_{j'})}{\theta_\lambda(t_j) \theta_\lambda(t_{j'})} 
\end{align*}
where $\widetilde{Y}_k(t) = \sum_{j=1}^d  w_j(t) Y_{k}(t_j)$ is the smoothed  trajectory for unit $k$, when there is no non response.
If we consider instead the ratio point of view, we have
\begin{align*}
\Var(\widehat{\mu}_{\lambda,Ha}^{(1)}(t)) & \approx
\left( 1 - \frac{n_\lambda}{N_\lambda} \right) \frac{1}{n_\lambda} \frac{1}{N_\lambda - 1} \sum_{k \in U_\lambda} \left( \widetilde Y_{k}(t) - \widetilde{\mu}_\lambda(t) \right)^2 \nonumber \\
 &+\frac{N_\lambda}{n_\lambda}\sum_{k\in U_\lambda}\sum_{j,j'=1}^d w_j(t) w_{j'}(t) (Y_k(t_j)-\widetilde{\mu}_\lambda(t))(Y_k(t_{j'})-\widetilde{\mu}_\lambda(t)) \frac{\theta_\lambda(t_j,t_{j'}) - \theta_\lambda(t_j) \theta_\lambda(t_{j'})}{\theta_\lambda(t_j) \theta_\lambda(t_{j'})}
\end{align*} 
and 
\begin{align*}
\Var(\widehat{\mu}_{\lambda,Ha}^{(2)}(t)) & \approx
\left( 1 - \frac{n_\lambda}{N_\lambda} \right) \frac{1}{n_\lambda} \frac{1}{N_\lambda - 1} \sum_{k \in U_\lambda} \left( \widetilde Y_{k}(t) - \widetilde{\mu}_\lambda(t) \right)^2 \nonumber \\
 &+\frac{N_\lambda}{n_\lambda}\sum_{k\in U_\lambda}\sum_{j,j'=1}^d w_j(t) w_{j'}(t) (Y_k(t_j)-\mu_\lambda(t_j))(Y_k(t_{j'})-\mu_\lambda(t_j)) \frac{\theta_\lambda(t_j,t_{j'}) - \theta_\lambda(t_{j'}) \theta_\lambda(t_{j'})}{\theta_\lambda(t_j) \theta_\lambda(t_{j'})}.
\end{align*} 
Note that, as expected, the part of the variance due to the sampling error is the same for three estimators, since they coincide when no data are missing.

\subsection{A comparison of the variances}
Defining \[
\Delta_\lambda(j,j') \ = \ \frac{\theta_\lambda(t_j,t_{j'}) - \theta_\lambda(t_j) \theta_\lambda(t_{j'})}{\theta_\lambda(t_j) \theta_\lambda(t_{j'})},
\]
the difference between the variances of the estimators is approximated as follows: 
\begin{align*}
\Var(\widehat{\mu}_{\lambda,Ha}^{(1)}(t))- \Var(\widehat{\mu}_{\lambda,HT}(t))&\approx \frac{N_\lambda}{n_\lambda}   \sum_{j,j'=1}^d  \Delta_\lambda(j,j')  w_j(t) w_{j'}(t) \sum_{k\in U_\lambda}  \widetilde{\mu}_\lambda(t) (\widetilde{\mu}_\lambda(t)-Y_k(t_{j'})-Y_k(t_j))  \\
 &= \frac{N_\lambda^2}{n_\lambda} \widetilde{\mu}_\lambda(t) \sum_{j,j'=1}^d  \Delta_\lambda(j,j')  w_j(t) w_{j'}(t)    \left(\widetilde{\mu}_\lambda(t)) - \mu_\lambda(t_j)  - \mu_\lambda(t_{j'}) \right)  
%&= \frac{N_\lambda^2}{n_\lambda} \frac{\theta_\lambda(t_j,t_{j'}) - \theta_\lambda(t_j) \theta_\lambda(t_{j'})}{\theta_\lambda(t_j) \theta_\lambda(t_{j'})}\widetilde{\mu}_\lambda(t)  \left(\sum_{j}^d w_j(t))   (\widetilde{\mu}_\lambda(t)-2\widetilde{\mu}_\lambda(t) \right) \\
%&= -\frac{N_\lambda^2}{n_\lambda} \frac{\theta_\lambda(t_j,t_{j'}) - \theta_\lambda(t_j) \theta_\lambda(t_{j'})}{\theta_\lambda(t_j) \theta_\lambda(t_{j'})}\left(\widetilde{\mu}_\lambda(t)\right)^2\\
\end{align*}

Considering now the $d \times d$ matrix $\boldsymbol{\Delta}_\lambda$, with generic elements $\Delta_\lambda(j,j')$,  previous difference can be expressed as follows
\begin{align*}
\Var(\widehat{\mu}_{\lambda,Ha}^{(1)}(t))- \Var(\widehat{\mu}_{\lambda,HT}(t))&\approx \frac{N_\lambda^2}{n_\lambda} \left( \mathbf{w}(t) \tilde{\mu}_\lambda(t) \right)^T\boldsymbol{\Delta}_\lambda \left(\mathbf{w}(t) \tilde{\mu}_\lambda(t) - 2 \breve{\boldsymbol{\mu}}_\lambda(t) \right)
\end{align*}
where  $\mathbf{w}(t) = (w_1(t), \ldots, w_d(t))$ and $\breve{\boldsymbol{\mu}}_\lambda(t) = (w_1(t) \mu_\lambda(t_1), \ldots, w_d(t) \mu_\lambda(t_d))$.
Since the bandwidth $h$ is small, $w_j(t)$ is very small (and supposed to be negligible) if $t$ is not very close to $t_j$ and $w_j(t) \approx 1$ if $t \approx t_j$. Thus, we can make the following approximation:
\begin{align*}
\left(\mathbf{w}(t) \tilde{\mu}_\lambda(t) - 2 \breve{\boldsymbol{\mu}}_\lambda(t) \right) &\approx - \mathbf{w}(t) \tilde{\mu}_\lambda(t).
\end{align*} 
Remarking that matrix $\boldsymbol{\Delta}_\lambda$  is non negative (it is a covariance matrix), we finally obtain that
\begin{align*}
\Var(\widehat{\mu}_{\lambda,Ha}^{(1)}(t))- \Var(\widehat{\mu}_{\lambda,HT}(t))&\approx -  \left( \mathbf{w}(t) \tilde{\mu}_\lambda(t) \right)^T\boldsymbol{\Delta}_\lambda \left(\mathbf{w}(t) \tilde{\mu}_\lambda(t) \right) \\ 
 & \leq 0.
\end{align*}
The H\'ajek estimator $\widehat{\mu}_{\lambda,Ha}^{(1)}(t)$ seems to be preferable to the Horvitz-Thompson  estimator $\widehat{\mu}_{\lambda,HT}(t)$ since it has a smaller variance when the bandwidth value $h$ is small, which has been supposed before in order to have a negligible bias. 

%In real applications, we have often have that the probabilities of response for two instants and for the same individual are positively correlated  
%\[
%\frac{\theta_k(t_j,t_{j'}) - \theta_k(t_j) \theta_k(t_{j'})}{\theta_k(t_j) \theta_k(t_{j'})}>0.
%\]
%In that case, $\Var(\widehat{\mu}_{\lambda,Ha}^{(1)}(t)) < \Var(\widehat{\mu}_{\lambda,HT}(t))$ and 
%hence for stratified random samples, the H\'ajek estimator is more precise than the Horvitz-Thomson estimator all the more so as the probabilities of missingness for differents instants are correlated.

\section{Some comments about estimation}

We discuss in this section some strategies that can be employed to estimate the mean trajectories and the variance of the estimators in practice.

\subsection{Variance estimation}
For the H\'ajek type of estimators,  we need to estimate the values of the linearized variables in order to  build an estimator of the variance. 
We can consider for example the following variance estimator (see Ardilly \& Till\'e, 2006, Chapter 9) for $\Var \left( \widehat{\mu}_{r,Ha}^{(1)}(t) \right)$:
\begin{eqnarray*}
\widehat{\Var} \left( \widehat{\mu}_{r,Ha}^{(1)}(t) \right) &=&\sum_{k\in s}\sum_{l\in s} \frac{\Delta_{kl}}{\pi_{kl} \pi_k\pi_l}\widehat{\widetilde u}^{(1)}_k(t)\widehat{\widetilde u}^{(1)}_l(t)\nonumber\\
& & +\sum_{k\in s}\frac{1}{\pi_k}\sum_{j,j'=1}^p\widehat{u}^{(1)}_{kj}(t)\widehat{u}^{(1)}_{kj'}(t)\frac{(\theta_k(t_{j},t_{j'})-\theta_k(t_j)\theta_k(t_{j'}))}{\theta_k(t_j)\theta_k(t_{j'})}r_{k}(t_{j})r_{k}(t_{j'}),
\end{eqnarray*}
where 
\begin{align*}
\widehat{u}_{kj}^{(1)}(t) &=  \frac{1}{N} w_j(t) \left(Y_{k}(t_j)-\widehat{\mu}_{r,Ha}^{(1)}(t) \right)
\end{align*}
and 
\[
\widehat{\widetilde u}^{(1)}_k(t) \ = \ \frac{1}{N} \sum_{j=1}^d  w_j(t) \left(Y_{k}(t_j)-\widehat{\mu}_{r,Ha}^{(1)}(t) \right)\frac{r_k(t_j)}{\theta_k(t_j)}.
\]
The estimator $\widehat{\Var} \left( \widehat{\mu}_{r,Ha}^{(2)}(t) \right) $ is obtained in a similar way.

\subsection{Suggestions on how to select the bandwidth values}

As it is well known in nonparametric kernel regression, when having to analyze real data, the quality of a nonparametric estimator strongly depends on how the value of the bandwidth is chosen. For example, it is shown in a similar context with a small simulation study in Cardot \textit{et al.} (2013) that linear interpolation can outperform kernel smoothing, even if the noise level is rather high, if the value of the bandwidth is chosen by a classical cross-validation performed curve by curve.  This individual procedure leads to oversmoothing (see also Hart \& Werhly, 1993), so that the bias of the resulting mean estimator is much larger than its variance. 
As in Cardot, Degras \& Josserand (2013), we suggest to use a modified cross-validation in order to choose the value of the bandwidth. This modified criterion    takes account of the sampling design as well as the non response process, the bandwidth value is chosen to minimize 
\begin{align}
CV(h) &= \sum_{\lambda=1}^\Lambda \sum_{k \in s_\lambda} \frac{N_\lambda}{n_\lambda} \sum_{j=1}^{d} \frac{r_k(t_j)}{\theta_\lambda(t_j)} \left( Y_k(t_{j}) -\widehat{\mu}^{(-k)}(t_{j}) \right)^2
\end{align}
where $\widehat{\mu}^{(-k)}$ is the estimator of the mean trajectory $\widetilde{\mu}$ built without considering trajectory $Y_k$ in the sample $s$. 
%More precisely, for stratified sampling, we can consider the following cross-validated mean estimator. If $k \in U_{\lambda_0}$, then
%\begin{align}
%\widehat{\mu}(t)^{(-k)} &= \frac{1}{N-1} \sum_{\lambda \neq \lambda_0} N_\lambda \widehat{\mu}_\lambda + \frac{1}{N-1} \sum_{\ell \neq k \in s_{\lambda_0}}  
%\end{align}
Note that considering different smoothing parameters in each stratum may not be more effective since it can lead, as noted before, to oversmoothing. Indeed, the best approximations, in terms of mean squared errors, of the mean of each subpopulation, may not lead to the best estimator of the overall mean function.  

\subsection{Estimation of the probabilities of response $\theta_{\lambda}(t_j)$ and $\theta_{\lambda}(t_j,t_{j'})$}
In the general situation in which each unit $k$ in the population is driven by a specific non trivial non response mechanism, it will be almost impossible to estimate the probability of response. If we suppose that, in the homogeneous  response groups context, the units within each group obey the same response mechanism, we can estimate the probability of response within each group by the response rate. If besides, the groups and the strata $U_\lambda$ coincide, then 
\[
\widehat{\theta}_{\lambda}(t_j)= \frac{1}{n_\lambda} \sum_{k \in s_{\lambda}}r_{k}(t_j)
\]
and 
\[
\widehat{\theta}_{\lambda}(t_j,t_{j'}) = \frac{1}{n_\lambda} \sum_{k \in s_{\lambda}}r_{k}(t_j) r_{k}(t_{j'}).
\]
It is easy to verify that in this case,  the Horvitz-Thompson estimator and the H\'ajek-type estimators are equal since, for all $\lambda=1, \ldots, \Lambda,$ we have:
$$
\widehat{\mu}_{\lambda,HT}(t)=\widehat{\mu}^{(1)}_{\lambda,Ha}(t)=\widehat{\mu}^{(2)}_{\lambda,Ha}(t)=\sum_{j=1}^dw_j(t)\frac{\sum_{s_{\lambda}}Y_k(t_j)r_k(t_j)}{\sum_{s_{\lambda}}r_k(t_j)}, \quad t\in [0, T].
$$
So,  $\widehat{\mu}_{\lambda,HT}(t)$ is a smoothed estimator of  $\frac{\sum_{s_{\lambda}}Y_k(t_j)r_k(t_j)}{\sum_{s_{\lambda}}r_k(t_j)}$ which  represents the mean of $Y_k(t_j)$ recorded on the respondent set from $s_{\lambda}$ and then, 
$$
\widehat{\mu}_{HT}(t)=\sum_{\lambda=1}^{\Lambda}\frac{N_{\lambda}}{N}\widehat{\mu}_{\lambda,HT}(t)
$$
 is a smoothed estimator of poststratified estimators (see S\"arndal \textit{et al.}, 1992).

We may also suppose that the response process is second order stationary, that is to say $\widehat{\theta}_{\lambda}(t_j)$ does not depend on $t_j$ and $\widehat{\theta}_{\lambda}(t_j,t_{j'})$ only depends on $|t_j - t_{j'}|$, so that we would get the following estimators 
\[
\widehat{\theta}_\lambda= \frac{1}{n_\lambda}  \sum_{k \in s_\lambda}\sum_{j=1}^d r_{k}(t_j)
\]
and, for each pair $(t_j,t_{j'})$ such that $| t_j -t_{j'} | = \Delta_t$,
\[
 \widehat{\theta}_{\lambda}(\Delta_t) = \frac{1}{n_\lambda}  \sum_{k \in s_\lambda }\sum_{j,j'  |t_j-t_{j'}|=\Delta_t} r_{k}(t_j) r_{k}(t_{j'}).
\]

\par 
These estimations can be performed either directly on the dataset used for the  mean load curve estimation or on a previous and  larger one, provided that it was collected by meters with the same technical characteristics.

%\par 

%When plugging-in the estimators of the response probabilities in the variance formula obtained in previous section, it can be shown that we still have, under classical hypotheses on the inclusion probabilities and the moments of the variable of interest,  a consistent estimator for the variance. 
\begin{comment}
\section{Concluding remarks - (a re-ecrire)}
\par
This work is currently going on with the comparison of the kernel strategy, on real data, with other inference strategies such as  linear interpolation and linear interpolation of the variation around the mean as well as  nearest neighbor imputation (see Beaumont \& Bocci, 2009, Kim \textit{et al.} 2011 or Shao \& Wang 2008). 
Contrary to the kernel estimation, these imputation methods have the advantage of providing imputed values to "fill the gaps" in the curves, enabling analysis at the individual level on incomplete datasets.
Some first tests have shown that the performances of linear interpolation are very good for short missing values series (a few hours) but decline very quickly when the gaps become larger. The linear interpolation of the variation around the mean works better for medium-size gaps but still fail when the gap is longer than a day.
\par
An other interesting direction for future investigation could be to take measurement errors into account. \marginpar{a developper}
\end{comment}

%\section{Comparison of the precision of the estimator}

%%%%%%%%%%%%%%%%%%%%%%%%%%%%%%%%%%%%%%%%%%
%%%%%%%%%%%%%%%%%%%%%%%%%%%%%%%%%%%%%%%%%%
\section*{Appendix : technical details}
\begin{itemize}
\item[\textbf{A1}.] We suppose that function $\mu$ is $\beta$-H\"older. There is  $\beta \in ]0,1]$ and  a constant $C$ such that $\forall (t,u) \in [0,T], \ |\mu(t)-\mu(u)| \leq C |t-u|^{\beta}$.
\item[\textbf{A2}.] We suppose that kernel $K$ is a continuous positive function with bounded support $[-1,1]$. 
%\item[A2.] We suppose that $\int_{\mathbb{R}} K(t) dt = 1$.
\item[\textbf{A3}.] We assume that the instants $0=t_1<t_2 < \cdots <t_d=T$ are equidistant, $t_j=(j-1)/(d-1), \ j=1, \ldots, d$ and the bandwidth satisfies  $2h>T(d-1)^{-1}$.
\end{itemize}
Conditions \textbf{A1} and \textbf{A2} are classical hypotheses in non parametric regression. Assumption \textbf{A2} is satisfied  for example if $K$ is the Epanechnikov kernel. Condition \textbf{A3}  ensures that the grid of discretization points is fine enough and that the bandwidth $h$ is not too small so that the estimator is well defined.

Let us write $\widetilde{\mu}(t)-\mu(t)$ as follows
\begin{align}
\widetilde{\mu}(t)-\mu(t) &= \frac{ \displaystyle \frac{1}{hd} \sum_{j=1}^{d} K\left(\frac{t-t_{j}}{h}\right) \left(\mu(t_{j})-\mu(t) \right)}{ \displaystyle \frac{1}{hd} \sum_{j=1}^{d} K\left(\frac{t-t_{j}}{h}\right) }
\label{def:ratiokernel}
\end{align}

Since kernel $K$ has a bounded support, we have that $K\left(\frac{t_j-t}{h} \right) >0$ only if $t_j \in [t-h,t+h].$ Since by assumption the instants of observation are equidistant in $[0,T]$, \textit{i.e.} $t_j = T(j-1)/(p-1), \ j=1,\ldots, p$, there are at most $2h(d-1)/T$ values of $K\left(\frac{t_j-t}{h} \right)$ that are strictly positive and hypothesis \textbf{A3} prevents all the terms in (\ref{def:ratiokernel}) from being equal to zero. Since function $\mu$ is $\beta$-H\"older, we have
\begin{align*}
 \left|\frac{1}{d} \sum_{j=1}^d \frac{1}{h} K\left(\frac{t_j-t}{h} \right) \left(\mu(t_j) -\mu(t)\right) \right| &\leq \frac{1}{d} \sum_{t_j \in [t-h,t+h]} \frac{1}{h} K\left(\frac{t_j-t}{h} \right)\left|\mu(t_j) -\mu(t)\right| \\
 & \leq  C |2h|^\beta  \ \frac{1}{d}  \sum_{t_j \in [t-h,t+h]} \frac{1}{h} K\left(\frac{t_j-t}{h} \right)
 %- \int_{t-h}^{t+h} \frac{1}{h}  K\left(\frac{u-t}{h} \right) \mu(u) du \right| & \leq \frac{2h (d-1)}{d} 
\end{align*} 
By Riemann sum approximation and the fact that kernel $K$ is continuous, with compact support, we get that, as $d \to \infty$, 
\begin{align*}
\left|\frac{1}{d} \sum_{j=1}^d \frac{1}{h}  K\left(\frac{t_j-t}{h} \right) - \frac{1}{h}\int  K\left(\frac{u-t}{h} \right)du \right| & \to 0
\end{align*}
and by the change of variable $x=(u-t)/h$ we have $\frac{1}{h}\int_{\mathbb{R}}  K\left(\frac{u-t}{h} \right)du = \int_{\mathbb{R}} K(x)dx < + \infty$.
We have proved that the bound given in (\ref{approx:errormutilde}) is true.

\bigskip

\noindent {\large \textbf{Acknowledgement}}. We would like to thank an anonymous referee for  helpful and valuable comments.

\section*{Bibliography}

\begin{description}

\item Ardilly, P. and Till\'e, Y. (2006). {\em Sampling Methods: Exercice and Solutions}. Springer, New York. 

%\item Beaumont, J-F. and Bocci, C. (2009). Variance estimation when donor imputation is used to fill in missing values. {\em The Canadian Journal of Statistics}, {\bf 37}, 400-416.

\item Cardot, H. and Josserand, E. (2011). Horvitz-{T}hompson estimators for functional data: asymptotic
  confidence bands and optimal allocation for stratified sampling.
 {\em Biometrika}, {\bf 98}, 107-118.

\item Cardot, H., Degras, D. and Josserand, E. (2013a). Confidence bands for Horvitz-Thompson estimators using sampled noisy functional data. 
 {\em Bernoulli}, {\bf 19}, 2067-2097.

\item Cardot, H., Dessertaine, A., Goga, C., Josserand, E. and Lardin, P. (2013b). Comparison of different sample designs and construction of confidence bands to estimate the mean of functional data: An illustration on electricity consumption.  {\em Survey Methodology}, {\bf 39}, 283-301.

\item Cardot, H., Goga, C. and Lardin, P. (2013c). Uniform convergence and asymptotic confidence bands for model-assisted estimators of the mean of sampled functional data.
\newblock{\em Electronic Journal of Statistics}, \textbf{7}, 562-596.   
%\item Chen, J. and Shao, J. (2000). Nearest Neighbor Imputation for Survey Data. {\em Journal of Official Statistics}, {\bf 16}, 113-131.

%\item Chen, J. and Shao, J. (2001). Jackknife Variance Estimation for Nearest-Neighbor Imputation. {\em Journal of the American Statistical Association}, {\bf 96}, 260- 269.

\item Deville, J-C. (1999). Variance estimation for complex statistics and estimators: linearization and residual techniques.
\newblock {\em Survey Methodology}, {\bf 25}, 193-203.

%\item Deville, J-C., S\"arndal, C-E. (1994). Variance estimation for the regression imputed Horvitz-Thompson estimator. {\em Journal of Official Statistics}, {\bf 10}, 381-394.

\item Hall  P., Mueller  H.G. and Wang  J.L. (2006). Properties of principal component methods for functional and
  longitudinal data analysis, \textit{Annals of Statistics}, \textbf{34},  1493-1517.

\item Hart, J. D. (1997). {\em Nonparametric smoothing and lack-of-fit tests}.
Springer Series in Statistics, Springer-Verlag, New York.

\item Hart, J. D. and Wehrly, T. E. (1986).
Kernel regression estimation using repeated measurements data. 
{\em J. Amer. Statist. Assoc.} {\bf 81}, 1080-1088.

\item Hart, J. D. and Wehrly, T. E. (1993). 
\newblock Consistency of cross-validation when the data are curves. 
\newblock {\em Stoch. Proces. Applic.}, {\bf 45}, 351-361.

\item Haziza, D. (2009). Imputation and inference in the presence of missing data. {\em Handbook of Statist., 29A, Sample surveys: design, methods and applications}. Elsevier/North-Holland, Amsterdam, 215-246. 

\item H\'ajek, J. (1971). Comment on the paper by D. Basu. In \textit{Foundation of Statistical Inference (eds. V.P. Godambe and D.A. Sprott)}, p. 236. Toronto: Holt, Rinehart and Winston.

\item Horvitz, D. and Thompson, D. (1952). 
\newblock A generalization of sampling without replacement from a finite universe. 
\newblock {\em Journal of the American Statistical Association}, \textbf{47}, 663-685.

%\item Haziza, D. (2007). Variance estimation for a ratio in presence of imputed data. {\em Survey Methodology}, {\bf 33}, 159-166. 

%\item Kim, J-K., Fuller, W-A. and Bell, W-R. (2011). Variance estimation for nearest neighbor imputation for US census long form data. {\em The Annals of Applied Statistics}, {\bf  5}, 824-842.

%\item Ramsay, J.O. and Silverman, B.~W. (2005).  {\em Functional data analysis}. Springer-Verlag, New York, second edition.

\item S\"{a}rndal, C-E., Lundstr\"om, S. (2005). \newblock {\em Estimation in Surveys with Nonresponse}. John Wiley \& Sons Ltd, Chichester, UK.

\item S\"{a}rndal, C-E., Swensson, B.~and Wretman, J. (1992). \newblock {\em Model assisted survey sampling}.
Springer Series in Statistics. Springer-Verlag, New York.

%\item Shao, J. (2009). Nonparametric Variance Estimation for Nearest Neighbor Imputation. {\em Journal of Official Statistics}, {\bf 25}, 55-62.

%\item Shao, J., Wang, H. (2008). Confidence Intervals Based on Survey Data with Nearest Neighbor Imputation. {\em Statistica Sinica}, {\bf 18}, 281-297.

\item Staniswalis, J.G., Lee, J.J. (1998). Nonparametric Regression Analysis of Longitudinal Data. {\em J. Amer. Statist. Assoc.}, \textbf{93}, 1403-1418
\end{description}

\end{document}